\documentstyle[axodraw]{article}

\newcommand{\beq}{\begin{equation}}
\newcommand{\eeq}{\end{equation}}
\newcommand{\beqa}{\begin{eqnarray}}
\newcommand{\eeqa}{\end{eqnarray}}
\def\bea{\begin{eqnarray}}
\def\eea{\end{eqnarray}}

\def\lam{\lambda}

\def\lp{\lambda'}
\def\ra{\rightarrow}
\newcommand{\Rsl}{{\not \! \!{R}}}
\def\ifmath#1{\relax\ifmmode #1\else $#1$\fi}
\def\rt2{\sqrt 2}
\def\half{\ifmath{{\textstyle{1 \over 2}}}}

\def\eq#1{Eq.~(\ref{#1})}

\newcommand{\snu}{\tilde \nu}

\def\cw{c_W}
\def\sw{s_W}
\def\nn{\nonumber}
\def\rt2{\sqrt 2}
\newcommand{\photino}{{\tilde \gamma}}

\def\SM{standard model}

\begin{document}

\vspace*{-1in}
\renewcommand{\thefootnote}{\fnsymbol{footnote}}

\begin{flushright}
\texttt{hep-ph/0406168} 
\end{flushright}
\vskip 5pt
\begin{center}
{\Large{\bf Neutrino masses and $R$-parity violation}}
\vskip 25pt
{\sf Subhendu Rakshit 
\footnote{E-mail address: srakshit@physics.technion.ac.il}} 
 
\vskip 10pt
{\it Department of Physics, Technion--Israel Institute of
  Technology\\
Technion City, 32000 Haifa, Israel} \\

\vskip 20pt

{\bf Abstract}
\end{center}

\begin{quotation}
{\small We review different contributions to the neutrino masses in the
context of $R$-parity violating supersymmetry in a basis independent
manner. We comment on the generic spectrum expected in such a scenario
comparing different contributions.
}
\end{quotation}

\vskip 20pt  

\setcounter{footnote}{0}
\renewcommand{\thefootnote}{\arabic{footnote}}

\section{INTRODUCTION}

In the last few years, neutrino oscillation experiments have
established that neutrinos do have tiny masses\cite{rev,rev2}.
However, in the standard model~(SM), neutrinos are massless. Hence
these observations are evidences of beyond SM physics.  The observed
pattern of neutrino masses and mixing angles guides us in the search
for new physics options beyond the SM.

Neutrino mass can be of Dirac type that conserves total lepton
number and requires the presence of a right-handed neutrino, or it can
be of Majorana type, that violates lepton number by two units. In the
\SM, we do not have right-handed neutrinos and the lepton number is
conserved. Hence neutrinos are massless. So SM needs to be
augmented. However, to explain the smallness of neutrino masses, it is
generally likely that an extension of the SM is lepton number
violating.  Supersymmetry offers a natural way out to accommodate
lepton number violation and hence, non-zero neutrino masses.

Minimal supersymmetric standard model~(MSSM) is an attractive
candidate for physics beyond the \SM. It solves many theoretical
puzzles posed by the SM and one expects to find its signatures in the
forthcoming colliders. In the MSSM, $R$-parity conservation is
imposed ad hoc on the Lagrangian to enforce lepton and baryon number
conservations. As a result, like the SM, MSSM also fails to accommodate
non-zero neutrino masses. However, Majorana neutrino masses can
be generated if the requirement of $R$-parity conservation is relaxed.
It is then interesting to explore how well $R$-parity violating
supersymmetric models can explain the observed neutrino mass and
mixing pattern.

Results from the recent neutrino oscillation experiments have more or
less converged and a rough outline of the mixing pattern has
emerged. It seems that it is quite different from our experience with
the quark sector, where the mixing angles are very small. The data is
best explained with the following set of parameters\cite{SKatmo03}
that control neutrino oscillations:
\beqa \label{fit}
&& \Delta m_{23}^2 = 2.0 \times 10^{-3}~{\rm eV}^2, \qquad \Delta
m_{12}^2 = 7.2 \times 10^{-5}~{\rm eV}^2, \nonumber\\
&& \sin^2\theta_{23}=0.5, \qquad \sin^2\theta_{12}=0.3,
\qquad \sin^2\theta_{13}<0.074,
\eeqa
where $\Delta m_{ij}^2 \equiv m_i^2-m_j^2$ are the squared mass
differences and $\theta_{ij}$ are the leptonic mixing angles.  It
suggests that there are one near-maximal ($\theta_{23}\sim 45^\circ$),
one large ($\theta_{12}\sim 30^\circ$), and one rather small
($\theta_{13}\le 15^\circ$) mixing angles. However, neutrino
oscillation experiments do not provide the overall mass scale. An
upper limit can be obtained from the recent WMAP results\cite{wmap} on
the cosmic microwave background radiation, which imply that for three
degenerate neutrinos, the common neutrino mass should be less than
$0.23$ eV. If we assume the lightest neutrino mass ($m_1$) to be
almost negligible, the data can be interpreted in a somewhat
``hierarchical'' mass scheme with $m_3\sim 0.04$eV and $m_2\sim
0.008$eV.

It is generally difficult to accommodate large mixing angles with
large mass hierarchy in a theoretical model. It can be
illustrated\cite{rev} with a two generation Majorana mass matrix
\beq
m_{\nu}=\pmatrix{a &b\cr b &c}.
\eeq
It is clear that one can get a large mass hierarchy but not the large
mixing when $a\gg b,c$ or one can get large mixing without a hierarchy
when $a,b,c\sim 1$ and $det(m_{\nu})\sim 1$. But one can get large
hierarchy with large mixing if $a,b,c\sim 1$ and $det(m_{\nu})\ll
1$. However, in general, this is not expected naturally. An underlying
mechanism is required to make the determinant small.  In addition, a
hierarchy in neutrino masses may be explained if different neutrinos
get masses from different sources. $R$-parity violating ($\Rsl$)
supersymmetric models are suitable in these regards.  In these models,
the heaviest neutrino gets mass at the tree level and the rest from
the loops, thus generating a hierarchy.

In this short review, rather than attempting to present a
comprehensive review of the entire literature\cite{newrev}, we will
give a pedagogical introduction\cite{valle} to the different
contributions to the neutrino masses in the MSSM with explicit
$R$-parity violation. We will work in a generic scenario, where, in
addition to the usual $R$-parity conserving couplings, all the lepton
number violating couplings allowed by the symmetries are present. In
the next section, we will describe our model and briefly discuss how
to express the results in a basis independent framework. In
section~\ref{sec:diagrams}, we will discuss different tree and loop
level diagrams contributing to the neutrino masses in detail. Their
contributions will be cast in a basis independent notation and
different suppression factors associated with them will be mentioned
in section~\ref{sec:generic}. We shall conclude in
section~\ref{sec:conclu}.

\section{$R\,$-PARITY VIOLATION}
In the SM, the field content and the requirement of renormalizability
lead to global symmetries of the Lagrangian ensuring lepton
number ($L$) and baryon number ($B$) conservations. In the MSSM, these
accidental symmetries are no longer present. $L$ and $B$ conservations
are guaranteed by imposing a discrete multiplicative symmetry called
`$R$-parity'\cite{Fayet:1974pd} on the Lagrangian. It is defined
as $R=(-1)^{3B+L+2S}$, so that for all SM particles $R=+1$ and for the
super-particles $R=-1$. $S$ denotes the spin of the particle.  However
in view of the observed proton stability, it is problematic to allow
both $L$ and $B$ violating terms simultaneously. Conservation of $B$
can be re-enforced by imposing a discrete baryon ${\mathbf Z}_3$
symmetry\cite{Ibanez:1991pr,Herbi}, thus allowing only $L$ violating
terms in the Lagrangian.

Once lepton number violation is allowed, there is no conserved quantum
number that distinguishes the lepton supermultiplets $\hat L_m$
($m=1,2,3$) from the down-type Higgs supermultiplet $\hat H_D$.  As a
consequence, the down-type Higgs (Higgsino) mixes with the sleptons
(leptons). It is therefore convenient to denote the four
supermultiplets by one symbol $\hat L_\alpha$ ($\alpha=0,1,2,3$), with
$\hat L_0\equiv \hat H_D$. We use Greek indices to indicate the four
dimensional extended lepton flavor space, and Latin ones for the usual
three dimensional flavor space.

The most general renormalizable lepton number violating superpotential
is given by\cite{ghrpv}:
\beq \label{rpvsuppot}
W=\epsilon_{ij} \left[
-\mu_\alpha \hat L_\alpha^i \hat H_U^j +
\half\lam_{\alpha\beta m}\hat L_\alpha^i \hat L_\beta^j \hat E_m^c +
\lp_{\alpha nm} \hat L_\alpha^i \hat Q_n^j  \hat D_m^c
-h^U_{nm}\hat H_U^i \hat Q^j_n \hat U_m^c
\right]\,,
\eeq
where $\hat H_U$ is the up-type Higgs supermultiplet; $\hat Q_n$ is a
doublet quark supermultiplet; $\hat U_m$, $\hat D_m$ and $\hat E_m$
are singlet up-type quark, down-type quark and charged lepton
supermultiplets respectively.  $\lambda_{\alpha\beta m}$ is
anti-symmetric under the interchange of the indices $\alpha$ and
$\beta$. Note that the $\mu$-term of the MSSM ($\mu_0$
in~\eq{rpvsuppot}) is now extended to a four-component vector
$\mu_\alpha$. The down-type quark (charged lepton) Yukawa matrix of the
MSSM, which corresponds to $\lp_{0ij}$ ($\lam_{0ij}$)
in~\eq{rpvsuppot}, is now extended to $\lp_{\alpha nm}$
($\lam_{\alpha\beta m}$).

One needs to include the possible $R$-parity violating terms in the
soft-supersymmetry-breaking sector as well. For example, one has to
add $R$-parity violating $A_{ijk}$, $A'_{ijk}$ and $B_i$ terms
corresponding to the superpotential $\Rsl$ terms $\lam_{ijk}$,
$\lp_{ijk}$ and $\mu_i$ respectively.  $\Rsl$ scalar squared-mass
terms also exist. The most general renormalizable $R$-parity violating
soft-supersymmetry-breaking potential is given by\cite{ghrpv}:
\bea \label{softsusy}
 V_{\rm soft}  &=&  (M^2_{\widetilde Q})_{mn}\,\widetilde Q^{i*}_m
        \widetilde Q^i_n
   +  (M^2_{\widetilde U})_{mn}\,\widetilde U_m^*\widetilde U_n
   +  (M^2_{\widetilde D})_{mn}\,\widetilde D_m^*\widetilde D_n
            \nn \\
 && +  (M^2_{\widetilde L})_{\alpha\beta}\,
          \widetilde L^{i*}_\alpha\widetilde L^i_\beta
      + (M^2_{\widetilde E})_{mn}\,\widetilde E_m^*\widetilde E_n
       + m^2_U|H_U|^2
  -(\epsilon_{ij} b_\alpha\widetilde L_\alpha^i H_U^j +{\rm h.c.}) \nn\\
 && +  \epsilon_{ij} \bigl[\half A_{\alpha\beta m} \widetilde L^i_\alpha
       \widetilde L^j_\beta \widetilde E_m + A'_{\alpha nm}
       \widetilde L^i_\alpha\widetilde Q^j_n\widetilde D_m
       - A^U_{nm} H^i_U
        \widetilde Q^j_n\widetilde U_m + {\rm h.c.}\bigr]\nn \\
 && +  \half \left[ M_3\, \widetilde g
   \,\widetilde g + M_2 \widetilde W^a\widetilde W^a
  + M_1 \widetilde B \widetilde B +{\rm h.c.}\right]\,.
\eea
The $B$-term of the MSSM is now extended to $B_\alpha$ in analogy with
$\mu_\alpha$ in the superpotential. $A$ and $A'$ terms are also
extended in a similar fashion. The squared-mass term for the down-type
Higgs boson and the $3\times 3$ slepton squared-mass matrix are
combined and extended to the $4\times 4$ matrix $(M^2_{\widetilde
L})_{\alpha\beta}$.

The total scalar potential comprises of contributions from the F-terms
as calculated from the superpotential~(\ref{rpvsuppot}), the soft
terms~(\ref{softsusy}), and the D-terms. The expressions can be found
in Ref.~\cite{ghrpv}. The contribution of the neutral scalar fields
to the scalar potential is given by,
\bea \label{Vn}
V_{\rm neutral} &=&
\left(m_U^2+ |\mu|^2\right) {|H_U|}^2 +
\left[({M^2_{\tilde L}})_{\alpha\beta} + \mu_\alpha\mu_\beta^*\right]
\snu_\alpha \snu_\beta^{*}\nn\\
&&
- \left(B_\alpha\snu_\alpha H_U +B_\alpha^*
\snu_\alpha^* H_U^*\right)
+\frac{1}{8}(g^2+g'^2)
\left[|H_U|^2-|\snu_\alpha|^2 \right]^2\,,
\eea
where $|\mu|^2\equiv \sum_\alpha|\mu_\alpha|^2$. Note that to simplify
notation, here we denote the neutral components of the up-type
scalar doublets by $H_U$. The vacuum expectation values (VEV)
for the neutral scalars denoted by $\langle H_U\rangle\equiv
\frac{v_u}{\rt2}$ and $\langle\snu_\alpha\rangle\equiv
\frac{v_\alpha}{\rt2}$ are determined by the following minimization
conditions which follow from \eq{Vn}:
\bea
(m_U^2+|\mu|^2)v_u^*
&=& B_\alpha v_\alpha-\frac{1}{8}(g^2+g^{\prime 2})
(|v_u|^2-|v_d|^2)v_u^*\,,\label{mincnda} \\
\left[({M^2_{\tilde L}})_{\alpha\beta} + \mu_\alpha\mu_\beta^*\right]
v_\beta^*
&=& B_\alpha v_u+\frac{1}{8}(g^2+g'^2)(|v_u|^2-|v_d|^2)v_\alpha^*\,,\label{mincndb}
\eea
with $|v_d|^2\equiv \sum_\alpha |v_\alpha|^2$.
We further define,
\beq
v\equiv (|v_u|^2+|v_d|^2)^{1/2}={2\,m_W\over g}\simeq 246~{\rm GeV},
\qquad \tan\beta \equiv \frac{v_u}{v_d}.
\eeq
For simplicity, we will assume that the gaugino mass parameters $M_i$,
$({M^2_{\tilde L}})_{\alpha\beta}$, $\mu_\alpha$, $B_\alpha$ and
$v_\alpha$ are real and choose $\tan\beta$ to be positive.

Treating $\hat{H}_d$ and $\hat{L}_i$ in the same footing entails a
freedom in choosing the direction of Higgs in the four dimensional
extended flavor space, which results from the absence of an unique
interaction eigenstate basis. However, talking in terms of the
numerical sizes of these $\Rsl$ couplings in the context of some
phenomenological consideration makes sense only if one specifies the
basis.  So one can either choose some specific basis; for example, one
can rotate to a basis where $\mu_i=0$ or $v_i=0$. The other
option\cite{sachaellis,Ferr,ghbasis} is to express experimental
observables in terms of basis invariant quantities, which we outline
in the following section.

\section{Basis Independence}
In the four dimensional vector space spanned by $(\hat{H}_d,
\hat{L}_1, \hat{L}_2, \hat{L}_3)$, one can construct\footnote{Here we
will follow Ref.~\cite{sachaellis}. For an alternative approach see
Ref.~\cite{ghbasis}.} coupling constant combinations invariant
under rotations in this space. It is interesting to note that it is
convenient to choose a basis while defining basis invariants. As a
first step one should pick up a Higgs direction. Although any
direction can be identified as the Higgs direction, the coupling
constants offer many directions to choose from. One can identify
$\mu_\alpha$ as the direction for the Higgs, for which
$\mu_i=0$. Similarly, $v_\alpha$, $\lp_{\alpha nm} \left[\equiv
(\lp_{nm})_\alpha\right]$ are vectors in this space and can very well
be chosen as Higgs directions. $\lam_{\alpha\beta p}$ are
anti-symmetric matrices and it helps us to choose the lepton
directions orthogonal to the Higgs direction, thus defining the
basis. An example will help to clarify the mechanism.

Let us choose the Higgs direction to be $v_\alpha/v_d$, which makes
sneutrino VEV's to vanish. The lepton directions are taken as $v^\alpha
\lam_{\alpha\beta p}/|v^\alpha \lam_{\alpha\beta p}|$. The anti-symmetry
in $\lam_{\alpha\beta p}$ ensures that the lepton directions are
orthogonal to the Higgs direction. The orthogonality amongst the
lepton directions is guaranteed by choosing the right-handed lepton
basis, such that $v^\alpha \lam_{\alpha\gamma p} \lam_{\beta\gamma q}
v^{\beta}\propto \delta_{pq}$. If $R$-parity is conserved, this
amounts to choosing the charged lepton mass eigenstate
basis. Similarly the left-handed quark and right-handed down-type
quark bases are chosen such that the down-type quark mass matrix
(proportional to $v^\alpha\lp_{\alpha pq}$) is diagonal. We have
neglected in this discussion the mixing between the charged leptons
and the charginos, induced by the $R$-parity violating bilinear
$\mu_i$ parameters, which are expected to be small due to the observed
smallness of the neutrino masses. These effects are included as
perturbations -- through ``mass insertions'' (explained afterwards) in
the Feynman diagram calculations, so that the results are basis
independent up to the order of square of the $\Rsl$ parameters. The
deviation of the charged lepton basis from its mass eigenstate basis
due to the presence of the small $\Rsl$ parameters will show up only
at higher orders.

We are now ready to write down the invariants with respect to the basis
we chose in the last paragraph. First we note that any scalar product
constructed out of the aforesaid vectors and matrices in this four
dimensional vector space are invariants. But we should choose those
which will be useful to express different contributions to neutrino
mass discussed in the next section. One such a set is given by\cite{Dav-Los}:
\bea
&&\delta^{\mu}_i = \frac{\mu^\beta}{|\mu|}\, \frac{v^{\alpha}\,\lam_{\alpha\beta i}}
{|v^\alpha\, \lam_{\alpha\beta i}|},
\quad~\qquad
\delta^{B}_i = \frac{B^\beta}{|B|} \,
\frac{v^{\alpha}\,\lam_{\alpha\beta i}}
{|v^{\alpha}\,\lam_{\alpha\beta i}|},\nn\\
&&\delta^{\lp}_{ipq} = 
\lp_{\beta pq}\, \frac{v^{\alpha}\,\lam_{\alpha\beta i}}
{|v^{\alpha}\,\lam_{\alpha\beta i}|},
\qquad
\delta^{\lam}_{ijk} = \frac
{v^\alpha\, \lam_{\alpha\beta i}}{|v^\alpha\, \lam_{\alpha\beta i}|}\,
\lam_{\beta\gamma k}\, 
\frac{v^{\theta}\,\lam_{\theta\gamma j}}
{|v^\theta \, \lam_{\theta\gamma j}|}\,.
\eea
$\delta^{\mu}_i$'s correspond to the projections of the vector
$\mu_\alpha/|\mu|$ onto the lepton directions. Similar geometric
interpretations can be made for the other invariants as
well\cite{sachaellis}.  These dimensionless invariants have the
property that they take simple forms in the vanishing sneutrino VEV
basis, $v_i=0$:
\beq\label{basis-inv}
\delta^{\mu}_i = \frac{\mu_i}{|\mu|},
\qquad
\delta^{B}_i = \frac{B_i}{|B|},
\qquad
\delta^{\lp}_{ijk} = \lp_{ijk},
\qquad
\delta^{\lam}_{ipq} = \lam_{ipq}\,.
\eeq
As the name suggests, the numerical values of these invariants remain
the same whichever basis one chooses, although their analytical forms
might change. In particular, they are zero if there is no $R$-parity
violation. Although these invariants were constructed choosing a
specific Higgs direction, one does not need to refer to any specific
basis while calculating experimental observables in terms of these
invariants, which is a must if one uses the usual coupling constants
instead of these invariant combinations. However, the simple forms
noted in \eq{basis-inv} suggest that one can proceed as follows,
rather than working in a general basis:
\begin{itemize}
 \item Work in the $v_i=0$ basis,
 \item calculate the neutrino mass
contributions in terms of the $\Rsl$ parameters $\mu_i$, $B_i$,
$\lam_{ipq}$, and $\lp_{ijk}$,
 \item and then replace these with
the basis invariant parameters $\delta^{\mu}_i$, $\delta^{B}_i$,
$\delta^{\lam}_{ipq}$, and $\delta^{\lp}_{ijk}$ respectively,
using \eq{basis-inv}.
\end{itemize}
This is equivalent to working in a general basis but is much
simpler. Thus one can get basis invariant results although in the
intermediate steps one can afford to work in a specific basis.  Here
we will generally work in $v_i=0$ basis. As we will see in the next
section that tree level neutrino contribution is unacceptably large
unless the vectors $v_\alpha$ and $\mu_\alpha$ are nearly parallel, we
expect that $\mu_i$ will be very small in this basis. This also means
that we will be very close to the charged lepton mass eigenstate
basis. The fact that the $\Rsl$ couplings are small in this basis will
also enable us to work in the ``mass insertion approximation'' which
means for propagating fields in the diagrams we will be using the
$R$-parity conserving usual MSSM mass eigenstates and include the
$R$-parity violating parameters as insertions in the diagram. Smaller
$\Rsl$ parameters would ensure better accuracy as we make this
approximation in the calculation of the Feynman diagrams in the next
section.

\section{CONTRIBUTIONS TO THE NEUTRINO MASSES}\label{sec:diagrams}
$R$-parity violating MSSM allows terms in the Lagrangian which
violates lepton number by one unit. Now two of these can be taken
together to construct a neutrino Majorana mass term which violates
lepton number by two units. Contributions to neutrino masses can
come from both tree and loop level diagrams and an extensive
literature exists on this subject\cite{all}. The one-loop
contributions from the trilinear $\Rsl$ couplings were much discussed,
until it was realized\cite{ghprl,HKK} that sneutrino--anti-sneutrino
mixing can play a significant role in generating neutrino masses at
one-loop. In Ref.~\cite{Dav-Los} many more loops were
identified. These loops were also studied in
Refs.~\cite{Davidson:1999mc,biloop,ygsr}. In what follows, we will
discuss only the dominant contributions in a generic scenario.

\subsection{TREE LEVEL CONTRIBUTION}
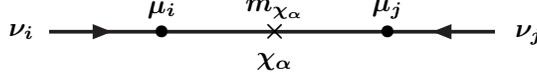
\begin{figure}[t]
\unitlength1mm
\SetScale{2.8}
\begin{boldmath}
\begin{center}
\begin{picture}(60,10)(0,-2)
\Line(0,0)(60,0)
\Text(-2,0)[r]{$\nu_i$}
\Text(62,-0.4)[l]{$\nu_j$}
\Text(30,-4)[c]{$\chi_{\alpha}$}
\Text(30,3)[c]{$m_{\chi_{\alpha}}$}
\Text(30,0)[c]{$\times$}
\Text(15,0)[c]{$\bullet$}
\Text(45,0)[c]{$\bullet$}
\Text(15,3)[c]{$\mu_i$}
\Text(45,3)[c]{$\mu_j$}
\SetScale{1.6}
\ArrowLine(0,0)(25,0)
\ArrowLine(100,0)(85,0)
\end{picture}
\end{center}
\end{boldmath}
\caption{Tree level neutrino mass in the mass
insertion approximation. A blob represents mixing between the neutrino
and the up-type Higgsino. The cross on the neutralino propagator
signifies a Majorana mass term for the neutralino.}
\protect\label{fig:tree-mass}
\end{figure}

As mentioned earlier, the neutrinos mix with neutralinos as $R$-parity
is violated. Hence, at the tree level, the neutrino mass matrix
receives contributions as shown in Fig.~\ref{fig:tree-mass}. In other
words,
the usual $4\times 4$ neutralino mass matrix in MSSM gets extended to
a $7\times 7$ neutralino--neutrino mass matrix. In $\{\widetilde
B,\widetilde W_3,\widetilde H_U, \widetilde H_D,\nu_1,\nu_2,\nu_3 \}$
basis, it is given by\cite{bgnn,enrico,ghrpv}:
\beq
M^{(n)}=\pmatrix{
M_1&0&m_Z\sw \frac{v_u}{v}&-m_Z\sw \frac{v_d}{v}& 0 & 0 & 0\cr
0&M_2&-m_Z\cw \frac{v_u}{v}&m_Z\cw \frac{v_d}{v}& 0 & 0 & 0\cr
m_Z\sw \frac{v_u}{v}&-m_Z\cw \frac{v_u}{v}&0&\mu&~\mu_1  & ~\mu_2 & ~\mu_3\cr
-m_Z\sw \frac{v_d}{v}&m_Z\cw \frac{v_d}{v}&\mu&0 & 0 & 0& 0\cr
0 & 0 &\mu_1&0 & 0 & 0& 0\cr
0 & 0 &\mu_2&0 & 0 & 0& 0\cr
0 & 0 &\mu_3&0 & 0 & 0& 0}\,
\eeq
where $\cw\equiv\cos\theta_W$ and $\sw\equiv\sin\theta_W$.  This is a
rank 5 matrix and taking into account all the four non-zero neutralino
masses, we see that only one neutrino gets massive at the tree
level. The four neutralinos can be integrated out to find the neutrino
mass matrix
\beq\label{mumu}
[m_\nu]_{ij}^{(\mu\mu)} = C \mu_i \mu_j\,,
\eeq
where,
\beq
C = {m_Z^2 m_{\tilde \gamma}\cos^2\beta \over
\mu(m_Z^2 m_{\tilde \gamma}\sin 2\beta-M_1 M_2 \mu)}
\sim {\cos^2\beta\over \tilde m}\,,
\eeq
such that $m_{\tilde \gamma}\equiv \cw^2 M_1 + \sw^2 M_2$ and in the
last step we assume that all the relevant masses are at the
electroweak (or supersymmetry breaking) scale, $\tilde m$.  The only
non-zero eigenvalue of $[m_\nu]_{ij}^{(\mu\mu)}$ is given by,
\beq
m_3 = C\, (\mu_1^2 + \mu_2^2+ \mu_3^2)\,.
\label{tree}
\eeq
This is the mass of the only neutrino which gets massive at the tree
level. We see that it is proportional to the $R$-parity violating
quantity $\sum\mu_i^2$ and to $\cos^2\beta$. For large $\tan
\beta$ the latter is a suppression factor.

For the following discussion, it is convenient to cast \eq{tree} in
a basis-invariant form
\beq
m_{3}=\frac{m_Z^2\, \mu\, m_{\photino} \cos^2\beta \,\sin^2\xi}{m_Z^2 m_{\photino} \sin 2\beta - M_1 M_2 \mu} \,,
\eeq
where $\xi$ is a measure of the misalignment of $v_\alpha$ and
$\mu_\alpha$:
\beq
\cos\xi = \frac{\sum_{\alpha} v_{\alpha} \mu_{\alpha}}{v_d \mu}.
\eeq
In a generic supersymmetric model one does not expect any alignment
between $v_{\alpha}$ and $\mu_{\alpha}$ and this leads to an
unacceptably large neutrino mass. This poses a problem for this kind
of models as fine-tuning is needed to achieve an exact or approximate
alignment. Moreover, this is not renormalization group invariant. That
is, if we achieve an alignment at some particular scale by fine-tuning
our model parameters, at some other scale a misalignment is expected
to pop up due to running of the parameters.

However, if some underlying mechanism ensures $v_{\alpha}\propto
\mu_{\alpha}$, the predicted mass can be reconciled with experimental
observations. Here the following observation plays a significant role.

Sufficient conditions\footnote{For a more general condition see
Ref.~\cite{ghrpv}.} to achieve the required alignment
$v_{\alpha}\propto \mu_{\alpha}$ are:
\begin{itemize}
\item
$B_{\alpha}$ is aligned with $\mu_{\alpha}$:
\beq
\label{conda}
B_{\alpha}\propto \mu_{\alpha},
\eeq
\item
$\mu_{\alpha}$ is an eigenvector of $(M^2_{\widetilde L})_{\alpha\beta}$:
\beq
\label{condb}
(M^2_{\widetilde L})_{\alpha\beta}\,\mu_{\beta}={\tilde m}^2 \mu_{\alpha}.
\eeq
\end{itemize}
To show this, it is convenient to go to a basis in which
$(M^2_{\widetilde L})_{\alpha\beta}$ is diagonal and consequently
\eq{condb} implies that in this basis $\mu_{\alpha}$ has only one
non-zero component, say $\mu_{0}$. Then \eq{conda} ensures that
for $B_{\alpha}$, only $B_{0}$ is non-zero. Then, as in the $R$-parity
conserving case, from the minimum equations it follows trivially
that $v_{\alpha}$ has the only non-zero component $v_0$. This implies
$v_{\alpha}\propto \mu_{\alpha}$.

The required alignment can be ensured by approximately satisfying
Eqs.~(\ref{conda}) and (\ref{condb}) in the framework of horizontal
symmetries\cite{bgnn}. Another approach is to consider high scale
alignment models. In these models, universality conditions at some
unification scale are assumed in a way to ensure that
Eqs.~(\ref{conda}) and (\ref{condb}) are satisfied exactly at that
scale. Then as one runs down the energy scale, the parameters
$(M^2_{\widetilde L})_{\alpha\beta}$ and $B_{\alpha}$ evolve,
generating a misalignment\cite{enrico} between $v_{\alpha}$ and
$\mu_{\alpha}$. This running is governed by charged lepton or
down-type quark Yukawa couplings and to have a misalignment it is
necessary to have a non-zero $L$-violating $\lambda$ or $\lambda'$
coupling or a corresponding $A$ term. If there are no suppressions
coming from these $L$-violating terms, the neutrino mass is at least
$1$~MeV ($1$~GeV)\cite{enrico} if the unification scale is taken to be
of the order of $M_{\rm Planck}\sim 10^{19}$ GeV ($10^6$~GeV). Hence
one needs a further strong suppression from these $L$-violating terms
to get a realistic neutrino mass.

We conclude this section noting that in a generic supersymmetric model
with $R$-parity violation, one can not avoid one neutrino being
massive at the tree level.

\subsection{TRILINEAR ($\lam\lam$ AND $\lp\lp$) LOOPS}
\begin{figure}[tb]
\unitlength1mm
\SetScale{2.8}
\begin{boldmath}
\begin{center}
\begin{picture}(60,20)(0,-2)
\Line(0,0)(60,0)
\Text(15,-4)[c]{$\lambda'_{ilk}$}
\Text(45,-4)[c]{$\lambda'_{jkl}$}
\DashCArc(30,0)(15,0,180){1}
\Text(13,15)[l]{$\tilde{d}_{k_{R}}$}
\Text(18,3)[l]{$d_{l_{L}}$}
\Text(45,15)[r]{$\tilde{d}_{k_{L}}$}
\Text(42,3)[r]{$d_{l_{R}}$}
\Text(-2,0)[r]{$\nu_i$}
\Text(62,-0.4)[l]{$\nu_j$}
\Text(30,0)[c]{$\times$}
\Text(30,15)[c]{$\bullet$}
\SetScale{1.6}
\ArrowLine(0,0)(25,0)
\ArrowLine(100,0)(85,0)
\end{picture}
\end{center}
\end{boldmath}
\caption[a]{Trilinear loop contribution to the neutrino mass matrix.
The blob on the scalar line indicates mixing between the left-handed
and the right-handed squarks. A mass insertion on the internal quark
propagator is denoted by the cross.}
\label{trilinear}
\end{figure}
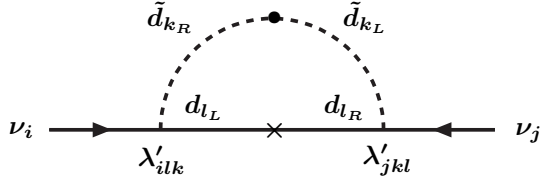

Neutrino masses receive contribution from fermion-sfermion
loops (see Fig.~\ref{trilinear}) dependent on the trilinear $R$-parity
violating couplings $\lam$ and $\lp$. These types of loops have
attracted much attention in the literature.  Here we only present
approximated expressions which are sufficient for our study. Complete
expressions can be found, for example, in Ref.~\cite{ghrpv}.

Neglecting quark flavor mixing, the contribution of the $\lp\lp$ loops
is proportional to the internal down-type quark mass and to the mixing
between left and right down-type squarks. Explicitly,
\beq
[m_\nu]_{ij}^{(\lp\lp)} \approx \sum_{l,k}\frac{3}{8\pi^2}
\lambda'_{ilk} \, \lambda'_{jkl} \,
\frac{m_{d_l}\, \Delta m_{\tilde d_k}^2}{m_{\tilde d_k}^2}
\sim
\sum_{l,k} \frac{3}{8\pi^2} \lambda'_{ilk} \lambda'_{jkl}
\frac{m_{d_l} m_{d_k}}{\tilde m},
\eeq
where $m_{\tilde d_k}$ is the average $k$th squark mass, $\Delta
m_{{\tilde d}_k}^2$ is the squared mass splitting between the two
$k$th squarks, and in the last step we used $\Delta m_{{\tilde
d}_k}^2\approx m_{d_k} \tilde{m}$ and $m_{\tilde
d_k}\sim\tilde{m}$. In the numerator $3$ is the color factor.  There
are similar contributions from loops with intermediate charged leptons
and sleptons where $\lambda'$ is replaced by $\lambda$ and there is no
color factor in the numerator:
\beq
[m_\nu]_{ij}^{(\lam\lam)} \approx \sum_{l,k}\frac{1}{8\pi^2}
\lambda_{ilk} \, \lambda_{jkl} \,
\frac{m_{\ell_l}\, \Delta m_{\tilde \ell_k}^2}{m_{\tilde \ell_k}^2}
\sim
\sum_{l,k} \frac{1}{8\pi^2} \lambda_{ilk} \lambda_{jkl}
\frac{m_{\ell_l} m_{\ell_k}}{\tilde m}.
\eeq
We see that the trilinear loop-generated masses are suppressed by the
$R$-parity violating parameters $\lambda^{\prime2}$ ($\lam^2$), by a
loop factor, and by two down-type quark (charged lepton) masses. The
latter factor, absent in other types of loops, makes the
trilinear contribution irrelevant in most cases.

\subsection{$BB$ LOOPS}
In the presence of bilinear $\Rsl$ terms, the sneutrinos and the
neutral Higgses mix at the tree level, which mixes sneutrinos and
anti-sneutrinos inducing a splitting between sneutrino mass
eigenstates. This generates a Majorana neutrino mass at one-loop. We
explain this mechanism in more detail in the following.

For simplicity let us work with one fermion generation. We assume CP
conservation. This enables us to consider CP even and CP odd scalar
sectors separately\cite{ghrpv,ghtalk}. The CP even sector consists of:
the light neutral Higgs $h$, the heavier neutral Higgs $H$, the CP
even sneutrino $\tilde\nu_+$. The CP odd scalar sector consists of:
the pseudoscalar Higgs $A$, the Goldstone boson $G$ (corresponding to
the $Z$), the CP odd sneutrino $\tilde\nu_-$. With $R$-parity
conservation, $\tilde\nu_{\pm}$ do not mix with the Higges and are
degenerate. So one defines the eigenstates of lepton number:
$\tilde\nu\equiv (\tilde\nu_{+}+i\,\tilde\nu_{-})/\sqrt 2$ and
$\bar{\tilde\nu}\equiv (\tilde\nu_{+}-i\,\tilde\nu_{-})/\sqrt 2$.  The
sneutrino mass squared term ($\Delta L=0$) is given by $m^2_{\snu}
\;\snu^* \snu$.

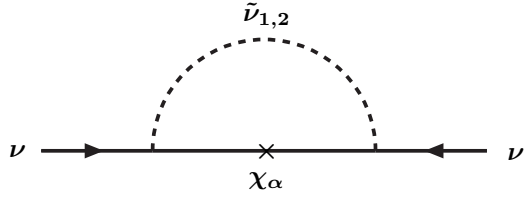
\begin{figure}[t]
\unitlength1mm
\SetScale{2.8}
\begin{boldmath}
\begin{center}
\begin{picture}(60,20)(0,-2)
\Line(0,0)(15,0)
\Line(45,0)(15,0)
\Line(60,0)(45,0)
\DashCArc(30,0)(15,0,180){1}
\Text(-2,0)[r]{$\nu$}
\Text(62,-0.4)[l]{$\nu$}
\Text(30,18)[c]{$\tilde\nu_{1,2}$}
\Text(30,-4)[c]{$\chi_{\alpha}$}
\Text(30,0)[c]{$\times$}
\SetScale{1.6}
\ArrowLine(0,0)(25,0)
\ArrowLine(100,0)(85,0)
\end{picture}
\end{center}
\end{boldmath}
\caption[a]{Majorana neutrino mass induced by sneutrino mass splitting.
The cross on the internal neutralino propagator denotes a Majorana
mass for the neutralino.}
\label{ghloop}
\end{figure}

When $R$-parity is violated, a $\Delta L=2$ mass term $m^2_{\Delta
L=2}\, \snu \,\snu$ gets generated which mixes the sneutrino and the
anti-sneutrino inducing a splitting between the sneutrino mass
eigenstates $\snu_{1,2}$. We assume the $\Rsl$ couplings to be small
which allows us to work perturbatively. Now $\tilde\nu_{+}$ mixes with
$h$ and $H$ and $\tilde\nu_{-}$ mixes with $A$ and $G$ and these
mixings are proportional to $B_1$. As a result, as the $3\times 3$
squared mass matrices in each CP sector gets diagonalized, the
sneutrino in each sector experiences a shift (to the leading order in
$B_1$) in their squared mass eigenvalues proportional to $B_1^2$. Now,
the two shifts are not equal, leaving behind a splitting between
$\snu_{1,2}$: $\Delta m_{\snu}\sim B_1^2/{\tilde m}^3$. It induces a
Majorana neutrino mass proportional to $\Delta m_{\snu}$ at one-loop
(see Fig.~\ref{ghloop}).

\begin{figure}[b]
\unitlength1mm
\SetScale{2.8}
\begin{boldmath}
\begin{center}
\begin{picture}(60,20)(0,-2)
\Line(0,0)(15,0)
\Line(45,0)(15,0)
\Line(60,0)(45,0)
\DashCArc(30,0)(15,0,180){1}
\Text(13,6)[c]{$\tilde\nu_i$}
\Text(20,11)[c]{$\bullet$}
\Text(16,14)[c]{$B_i$}
\Text(47,5.6)[c]{$\tilde\nu_j$}
\Text(40,11)[c]{$\bullet$}
\Text(43,13.6)[c]{$B_j$}
\Text(-2,0)[r]{$\nu_i$}
\Text(62,-0.4)[l]{$\nu_j$}
\Text(30,18)[c]{$h,H,A~$}
\Text(30,0)[c]{$\times$}
\Text(30,-4)[c]{$\chi_{\alpha}$}
\SetScale{1.6}
\ArrowLine(0,0)(25,0)
\ArrowLine(100,0)(85,0)
\end{picture}
\end{center}
\end{boldmath}
\caption[a]{The $BB$ loop-generated neutrino mass. This is the same diagram
as in Fig.~\ref{ghloop}, but re-drawn in the mass insertion
approximation.  Here the blobs denote mixing of the sneutrinos with
the neutral Higgs bosons.}
\label{fig-BBloop}
\end{figure}
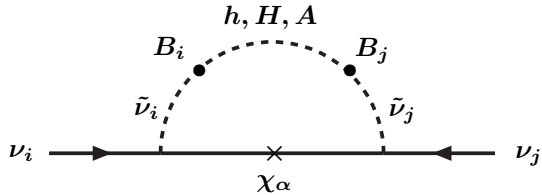

For quantitative estimations, we will recast the same diagram in the
``mass insertion approximation'' as shown in Fig.~\ref{fig-BBloop}.
The contribution from this diagram is proportional to two insertions
of $B_i$ parameters.
However, it should be mentioned that in addition to the $B_i$ terms,
the terms like $({M^2_{\tilde L}})_{0i}$ and $\mu_i$
also contribute (see \eq{Vn}). However, these contributions are
related to $B_i$ through the minimum equation (\eq{mincndb})
in the $v_i=0$ basis, namely $[({M^2_{\tilde L}})_{0i} +
\mu_0\mu_i] = -\tan\beta\,B_i$. So we will designate these
all-bilinear loops as $BB$ loops.

The contribution to the neutrino mass matrix from $BB$ loop is
given by\cite{Dav-Los},
\bea \label{BBloop}
[m_\nu]^{(BB)}_{ij}
& = & \sum_{{\alpha}}
\frac{g^2 B_i B_j}{4 \cos^2\beta}
(Z_{{\alpha}2} - Z_{{\alpha}1} g'/g)^2  m_{\chi_{\alpha}}
 \left\{ I_4(m_h, m_{\tilde{\nu}_i},m_{\tilde{\nu}_j},
m_{\chi_{\alpha}})
\cos^2(\alpha - \beta) \right. \nonumber \\
 && \left. +
I_4(m_H, m_{\tilde{\nu}_{i}}, m_{\tilde{\nu}_{j}},
m_{\chi_{\alpha}}) \sin^2(\alpha - \beta)
- I_4(m_A, m_{\tilde{\nu}_{i}},m_{\tilde{\nu}_{j}},
m_{\chi_{\alpha}}) \right\}\,,
 \label{GH}
\eea
where $Z_{\alpha \beta}$ is the neutralino mixing matrix with
$\alpha,\beta=1,..,4$ and
\beqa \label{Ifour}
I_4(m_{1}, m_{2}, m_{3}, m_4) &=& {1\over m_1^2 -m_2^2} \left[I_3(m_{1},
m_{3}, m_4)-I_3(m_{2}, m_{3}, m_4)\right],\nonumber \\
I_3(m_{1}, m_{2}, m_{3}) &=& {1\over m_1^2 -m_2^2}
\left[I_2(m_{1}, m_{3}) - I_2(m_{2}, m_{3})\right],\nonumber \\
I_2(m_{1}, m_{2 }) &=& - {1\over 16\pi^2} {m_1^2\over
m_1^2 -m_2^2} \ln\frac{m_1^2}{m_2^2}.
\eeqa
Assuming that all the masses in the right-hand side of \eq{BBloop} are
of the order of the weak scale, we estimate
\beq \label{BB-approx}
[m_\nu]_{ij}^{(BB)} \sim \frac{g^2}{64\pi^2}\,\frac{1}{\cos^2\beta}\,
{B_i B_j \over  \tilde{m}^3}\, \epsilon_H
\eeq
where,
\beq \label{def-eps-H}
\epsilon_{H} \equiv
\left|{I(m_h)\, \cos^2(\alpha - \beta)\, +\, I(m_H)\, \sin^2(\alpha - \beta)\, -\, I(m_A)
\over
|I(m_h)| \cos^2(\alpha - \beta) + |I(m_H)| \sin^2(\alpha - \beta) +
|I(m_A)|}\right|\,,
\eeq
where $I(x)\equiv I_4(x, m_{\tilde{\nu}_{i}}, m_{\tilde{\nu}_{j}},
m_{\chi_{\alpha}})$ [see \eq{BBloop}], and the $i,j,\alpha$ indices
of $\epsilon_H$ are implicit.

Naively, one would expect $\epsilon_{H}\sim 1$. However, it has been
pointed out in Ref.~\cite{ygsr} that it is parameter space
dependent and can provide a suppression of several orders of
magnitude.  This suppression is related to the Higgs decoupling which
is typical\cite{Haber} to the two Higgs doublet models like MSSM. At
the decoupling limit, $\cos^2(\alpha - \beta)\to 0$, $m_H\simeq m_A
\gg m_h \simeq m_Z$, we see that $\epsilon_H\ra 0$. The three
Higgs loops tend to cancel each other, which becomes more and more
severe as we approach the decoupling limit. This can be further
clarified if we look at the weighted sum of the three Higgs
propagators before integrating over the internal momenta $k$:
\beq
P_S=\frac{1}{k^2-m_h^2} \cos^2(\alpha-\beta) + \frac{1}{k^2-m_H^2}
\sin^2(\alpha-\beta) - \frac{1}{k^2-m_A^2}.
\eeq
For simplicity we use the tree level relations\cite{hunter,gh1}
\beq\label{htree}
\cos^2(\alpha-\beta)=\frac{m_h^2(m_Z^2-m_h^2)}{m_A^2(m_H^2-m_h^2)},\qquad
m_Z^2-m_h^2=m_H^2-m_A^2,
\eeq
to obtain
\beq \label{PSff}
P_S=\frac{-k^2(m_Z^2-m_h^2)(m_A^2-m_h^2)}
{m_A^2(k^2-m_H^2)(k^2-m_A^2)(k^2-m_h^2)}.
\eeq
In the approach to the decoupling limit, $P_S$ scales inversely to the
fourth power of the heavy mass. Thus the weighted sum falls faster
than the individual propagators contributing to $P_S$, indicating an
increasingly stronger cancellation between the Higgs diagrams. As one
includes loop corrections, modifying the tree level relations
\eq{htree}, some cancellation still occurs. The point to note is that
a partial cancellation occurs even away from the decoupling limit as
the $I_4$ functions do not change much with respect to its
arguments. This makes $\epsilon_H$ to emerge as a suppression factor.

Another interesting point is that this Higgs cancellation becomes more
significant for high $\tan\beta$. This significantly reduces the
enhancement effect from the $1/\cos^2\beta$ factor in \eq{BBloop} for
high $\tan\beta$, making $[m_\nu]^{(BB)}_{ij}$ less sensitive to
$\tan\beta$.

Next we study the $BB$ loop effect on the neutrino masses.  For this
we rewrite \eq{BBloop} as
\beq\label{BB}
[m_\nu]^{(BB)}_{ij}=C_{ij}\,B_i B_j.
\eeq
Now if sneutrinos of all generations are degenerate, all $C_{ij}$'s
will be equal and only one neutrino will be massive from the $BB$
loops. In general one does not expect $B_\alpha \propto
\mu_\alpha$. So the neutrino which gets mass from the $BB$ loops
is not the same one which got massive at the tree level (see
\eq{mumu}). Hence if we consider only the tree level and $BB$ loop
contributions, in the case of degenerate sneutrinos one neutrino is
left massless. Although it can get mass from other loops, it is
interesting to note that it can be massive from the $BB$ loops itself
if the sneutrinos are non-degenerate\footnote{It should be emphasized
that here we are talking about non-degeneracy of sneutrinos of
different generations, which is different from the non-degeneracy of
the sneutrino mass eigenstates of a given generation induced by the
bilinear $R$-parity violating parameters.}, making $C_{ij}$ no longer
a rank one matrix. Hence one would expect that the contribution to the
lightest neutrino mass from $BB$ loops will be dependent on the amount
of non-degeneracy in the sneutrino sector.  In what follows, we will
try to quantify this dependence when the non-degeneracy is
small compared to the sneutrino masses.

We assume that the heaviest neutrino acquires large mass at the tree
level. This helps us to deal only with the loop contributions to the
first two generations, simplifying our problem.

We take the mass-squares of the two sneutrinos as $(m^2_{\snu})_{1,2}
\equiv m_{\snu}^2 (1\pm\Delta)$. After we plug them in \eq{BBloop}
and diagonalize, we get the following mass eigenvalues in their
leading order:
\beq
m_{2} = (B_1^2+B_2^2)\, f_1 ,\qquad
m_{1} = \frac{B_1^2 B_2^2}{B_1^2+B_2^2} \,(4 f_1 f_3 - f_2^2)\,
\Delta^2 ,
\label{masses}
\eeq
where $f_i$'s are the coefficients\cite{ygsr} of the loop function
expanded in powers of $\Delta$.  We see that $m_{2}$ is the same as
that in the degenerate case. $m_{1}$ is proportional to the square of
the sneutrino mass splitting between generations.

We define the following measure of the degeneracy suppression
\beq \label{def-eps-D}
\epsilon_D\equiv {m_1 \over m_2}=f_c\,
\frac{B_1^2 B_2^2}{(B_1^2+B_2^2)^2}\, \Delta^2,
\qquad f_c= {4 f_1 f_3-f_2^2 \over f_1}.
\eeq
We note that, in addition to the $\Delta^2$ suppression, the lightest
neutrino mass is also suppressed by $f_c\sim 0.1$\cite{ygsr}.

\subsection{$\mu B$ LOOPS}
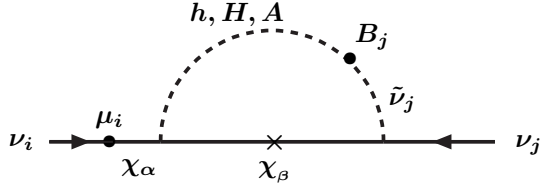
\begin{figure}[tb]
\unitlength1mm
\SetScale{2.8}
\begin{boldmath}
\begin{center}
\begin{picture}(60,20)(0,-2)
\Line(0,0)(15,0)
\Line(45,0)(15,0)
\Line(60,0)(45,0)
\DashCArc(30,0)(15,0,180){1}
\Text(8,0)[c]{$\bullet$}
\Text(8,3)[c]{$\mu_i$}
\Text(47,5.6)[c]{$\tilde\nu_j$}
\Text(40,11)[c]{$\bullet$}
\Text(43,14)[c]{$B_j$}
\Text(-2,0)[r]{$\nu_i$}
\Text(62,-0.4)[l]{$\nu_j$}
\Text(25,17)[c]{$h,H,A$}
\Text(30,0)[c]{$\times$}
\Text(12,-3.5)[c]{$\chi_\alpha$}
\Text(30,-4)[c]{$\chi_{_\beta}$}
\SetScale{1.6}
\ArrowLine(0,0)(15,0)
\ArrowLine(100,0)(85,0)
\end{picture}
\end{center}
\end{boldmath}
\caption[a]{
Neutrino Majorana mass generated by $\mu B$ loop.  The cross on the
internal neutralino line denotes, as before, a Majorana mass term for
the neutralino. There exists another diagram with $i\leftrightarrow
j$.}
\label{fig-muBloop}
\end{figure}

There is another type of diagram (Fig.~\ref{fig-muBloop}), which
involves both neutrino--Higgsino mixing (through $\mu_i$ terms) and
sneutrino--Higgs mixing (through $B_i$ terms).
The contribution from this diagram to the neutrino mass matrix is
given by\cite{Dav-Los,ygsr}\footnote{Note that in this expression,
there is a typographical error in sign in Ref.~\cite{Dav-Los},
which has been corrected in Ref.~\cite{ygsr}.},
\bea
[m_\nu]_{ij}^{(\mu B)}&=& \sum _{\alpha, \beta}
 {{g^2} \over  4 \cos \beta}
 \mu_{i}  B_j
{m_{\chi_{_\beta}} \over m_{\chi_{\alpha}}}
Z_{\alpha 3} ( Z_{\beta 2} -  Z_{\beta 1} g'/g)
\nonumber \\ &&
\bigg\{-\Big[
 Z_{\alpha 4}( Z_{\beta 2}-  Z_{\beta 1} g'/g) \sin \alpha +
   ( Z_{\alpha 2}-  Z_{\alpha 1} g'/g)Z_{\beta 3} \cos \alpha
\nonumber \\&& \quad\quad
+ ( Z_{\alpha 2}-  Z_{\alpha 1} g'/g)  Z_{\beta 4}\sin \alpha \Big]
  \cos(\alpha - \beta) \,
I_3(m_h,m_{\chi_{_\beta}}, m_{\tilde{\nu}_j})
\nonumber  \\ &&\quad
+ \Big[  Z_{\alpha 4}( Z_{\beta 2}-  Z_{\beta 1} g'/g) \cos \alpha
 -( Z_{\alpha 2}-  Z_{\alpha 1} g'/g)  Z_{\beta 3}\sin \alpha
\nonumber \\&& \quad\quad
+( Z_{\alpha 2}-  Z_{\alpha 1} g'/g) Z_{\beta 4} \cos \alpha \Big]
   \sin(\alpha - \beta) \,
I_3(m_H,m_{\chi_{_\beta}}, m_{\tilde{\nu}_j})
\nonumber \\ && \quad
+ \Big[ Z_{\alpha 4}( Z_{\beta 2}-  Z_{\beta 1} g'/g) \sin \beta
 + ( Z_{\alpha 2}-  Z_{\alpha 1} g'/g)Z_{\beta 3} \cos \beta
\nonumber \\&& \quad\quad
+ ( Z_{\alpha 2}-  Z_{\alpha 1} g'/g)  Z_{\beta 4}\sin \beta \Big] \,
I_3(m_A,m_{\chi_{_\beta}}, m_{\tilde{\nu}_j})
\bigg\}  + (i \leftrightarrow j)
\label{muB-term}
\eea
Taking all the weak scale masses as $\tilde m$, these contributions to
the neutrino mass matrix are given approximately by
\beq\label{muB-approx}
[m_\nu]_{ij}^{(\mu B)}\sim \frac{g^2}{64\pi^2}\, \frac{1}{\cos\beta}\,
\frac{\mu_i B_j+\mu_j B_i}{\tilde m^2}\, \epsilon'_H.
\eeq
Here $\epsilon'_H$ is a similar suppression factor expected from
cancellation between different Higgs diagrams as in the case of $BB$
loops.

Comparing \eq{muB-approx} with \eq{BB-approx}, we see that in the
flavor basis these diagrams are expected to yield similar
contributions to the $BB$ loops. But due to the dependence on $\mu_i$,
the $\mu B$ loop contribution to the neutrino masses is
sub-leading\cite{Chun,ygsr}, if the tree level contribution is the
dominant one.  For illustration, let us consider a two generation case
and consider contributions from the tree level and $\mu B$ diagrams
only. Now a look at \eq{muB-term} reveals that one can recast it in
the following form\cite{ygsr}:
\beq
[m_\nu]^{(\mu B)}_{ij} = C \varepsilon_L (\mu_i {\cal B}_j +
\mu_j {\cal B}_i)
\eeq
where $C$ is $i,j$ independent (complies with the definition in
\eq{mumu}), $\varepsilon_L$ is the loop suppression factor and ${\cal
B}_i$ corresponds to the product of $B_i$ and the loop function which
also carries the index $i$. Together with the tree level contribution
$[m_\nu]^{(\mu\mu)}_{ij} = C \mu_i \mu_j$, the mass matrix looks like
\beq
[m_\nu]^{(\mu\mu+\mu B)} = C
\pmatrix{
\mu_1^2 + 2\, \varepsilon_L {\cal B}_1 \mu_1 & \mu_1 \mu_2 + \varepsilon_L
({\cal B}_1\mu_2 +{\cal B}_2 \mu_1) \cr \mu_1 \mu_2 + \varepsilon_L
({\cal B}_1\mu_2 +{\cal B}_2 \mu_1) &\mu_2^2 + 2\, \varepsilon_L {\cal
B}_2 \mu_2}.
\eeq
After diagonalization, we see that the lighter eigenvalue is
suppressed with respect to the heavier one by a factor of the order of
$\varepsilon_L^2$. This suppression is like ${\cal O}(\varepsilon_L)$
for the $BB$ loops as in that case $C$ will dependent on $i,j$ (see
\eq{BB}).

Hence we conclude that although in the flavor basis $\mu B$ loops
seems to have comparable contributions to the $BB$ loops, in the
presence of large tree level contributions they contribute
insignificantly to the neutrino masses after diagonalization. We will
not elaborate on this diagram any further.

\subsection{$\mu\lam$ AND $\mu\lp$ LOOPS}

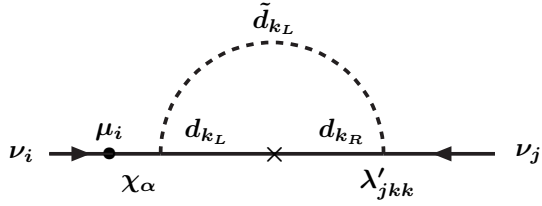
\begin{figure}[tb]
\unitlength1mm
\SetScale{2.8}
\begin{boldmath}
\begin{center}
\begin{picture}(60,20)(0,-2)
\Line(0,0)(15,0)
\Text(8,0)[c]{$\bullet$}
\Text(8,3)[c]{$\mu_i$}
\Line(45,0)(15,0)
\Line(60,0)(45,0)
\Text(45,-4)[c]{$\lambda'_{jkk}$}
\DashCArc(30,0)(15,0,180){1}
\Text(18,3)[l]{$d_{k_{L}}$}
\Text(30,18)[c]{$\tilde{d}_{k_{L}}$}
\Text(42,3)[r]{$d_{k_{R}}$}
\Text(-2,0)[r]{$\nu_i$}
\Text(62,0)[l]{$\nu_j$}
\Text(30,0)[c]{$\times$}
\Text(12,-4)[c]{$\chi_{\alpha}$}
\SetScale{1.6}
\ArrowLine(0,0)(15,0)
\ArrowLine(100,0)(85,0)
\end{picture}
\end{center}
\end{boldmath}
\caption[b]{
Neutrino Majorana mass generated by $\mu \lp$ loop.  There
exists another diagram with $i\leftrightarrow j$.}
\label{fig-mulam}
\end{figure}

There are contributions to the neutrino masses from loops containing
both bilinear ($\mu_i$) and trilinear ($\lam$ or $\lp$) $\Rsl$
couplings (see Fig.~\ref{fig-mulam}). The blob on the external line
requires some explanation. The neutrino gets converted to a up-type
Higgsino through $\mu_i$ terms. This Higgsino mixes with gauginos
which then couple with $\tilde{d}_{L}$ and $d_{L}$. The cross on the
internal fermion line represents a Dirac mass insertion. Neglecting
squark flavor mixing, the $\mu\lp$ contribution to the neutrino mass
matrix is given approximately by\cite{Dav-Los}
\beq
[m_\nu]_{ij}^{(\mu\lp)} \approx \sum_k \frac{3}{16 \pi^2}\, g\, m_{d_k}\,
 \frac{\mu_i \lambda'_{jkk}+\mu_j \lambda'_{ikk}}   {\tilde m}.
\eeq
A factor of $3$ appears in the numerator as a color factor. There are
similar diagrams comprising of $\lam$ instead of $\lp$ with a color
factor $1$, where the quarks and squarks are replaced by charged
leptons and sleptons respectively. If we compare these contributions
with $\lp\lp$ or $\lam\lam$ type diagrams which suffer from two Yukawa
suppressions, we see that this diagram has only one Yukawa
suppression.  But arguing in the same line as the $\mu B$ loops, one
can see that these diagrams do not contribute to the neutrino masses
significantly in the presence of large tree level contributions.

\section{DISCUSSIONS}\label{sec:generic}
In the previous section we have seen that the different contributions
to neutrino masses come with different suppression factors. In this
section, for completeness, we summarize these contributions in terms
of basis independent parameters following Eqs.~(\ref{basis-inv}) and
comment on their relative contributions.

Approximate expressions of different contributions are given by:
\bea
{[m_\nu]}_{ij}^{(\mu\mu)} &\sim& \delta^{\mu}_i \, \delta^{\mu}_j
\,\cos^2\beta\, {\tilde m}, \nn\\
{[m_{\nu}]}_{ij}^{(\lp\lp)} &\sim&
\sum_{l,k} \frac{3}{8\pi^2} \delta^{\lp}_{ilk} \, \delta^{\lp}_{jkl}\,
\left(\frac{m_{d_l} m_{d_k}}{{\tilde m}^2}\right)\,\tilde m ,\nn\\
{[m_\nu]}_{ij}^{(BB)} &\sim& \frac{g^2}{64\pi^2} \frac{\epsilon_H}{\cos^2\beta}
\delta^B_i \, \delta^B_j\, {\tilde m} , \\
{[m_{\nu}]}_{ij}^{(\mu B)} &\sim& \frac{g^2}{64\pi^2}\,
\frac{\epsilon'_H}{\cos\beta} (\delta^\mu_i\, \delta^B_j + \delta^\mu_j\,
\delta^B_i)\, {\tilde m} , \nn\\
{[m_{\nu}]}_{ij}^{(\mu\lp)} &\sim& \sum_k \frac{3}{16 \pi^2}\, g\,
(\delta^\mu_i \, \delta^{\lambda'}_{jkk}+\delta^\mu_j\,
\delta^{\lambda'}_{ikk})\,\left(\frac{m_{d_k}}{\tilde m}\right)\,\tilde m, \nn
\eea
where all the weak scale parameters are taken as $\tilde m$, which is
then factored out as an overall scale.

Although the leading effects are model dependent, some general
remarks can be made, if we assume that all the $\Rsl$ parameters
are of the similar order.
\begin{itemize}
\item
$m_3$ receives a contribution at the tree level and this is the
dominant one unless $\tan\beta$ is very large. $m_2$ and $m_1$ are
generated at one-loop. Apart from being down by a loop factor $\sim
\frac{1}{16\pi^2}$, the loops suffer from several other suppression factors.
\item
Although $BB$ loop can get a suppression $\epsilon_H$ from a partial
cancellation between different Higgs loops, unless one is too close to
the decoupling regime, it is generally likely that this is the most
dominant loop contribution. The leading contribution to $m_2$ comes
from this loop. A $\tan\beta$ dependence comes from $\epsilon_H$ which
partly cancels the $\cos^2\beta$ in the denominator reducing the
sensitivity to $\tan\beta$.
\item
The trilinear loops $\lp\lp$ and $\lam\lam$ suffer from double Yukawa
suppression, and contribute very little.
\item
Apart from a suppression $\epsilon'_H$ from a partial cancellation
between different Higgs loops, the $\mu B$ loop contribution to the
neutrino mass is suppressed in presence of unsuppressed tree level
contribution.
\item
The contributions from $\mu\lam$ and $\mu\lp$ loops also have a
similar suppression for large tree level contribution and are further
down by a single Yukawa suppression.
\item
Although $m_1$ can receive contributions from other loops, it can be
massive from $BB$ loops alone if the sneutrinos are non-degenerate. If
the amount of non-degeneracy is not small, the dominant contribution to
$m_1$ can very well come from $BB$ loops and it will be suppressed
with respect to $m_2$ by $\epsilon_D$. In this case, the tree level
and the $BB$ loops provide the most important contributions to all
neutrinos. Note that since neutrino oscillation data are not sensitive
to the overall scale of the neutrino masses, our ignorance of the
mechanism that generate $m_1$ is not problematic.
\end{itemize}

Now we can comment on how $R$-parity violating supersymmetric
models confront the experimental observations. We will estimate
the approximate sizes required to fit the data, when the tree
level contribution dominates over the loops. However, for detailed
numerical analysis see Refs.~\cite{Chun,numerical}.

If we neglect $m_1$, we can infer from \eq{fit},
\bea
m_3 &\sim& (\delta^{\mu})^2  \,\cos^2\beta\,
{\tilde m} \sim 10^{-1} eV\nn\\
m_2 &\sim& \frac{g^2}{64\pi^2} \frac{1}{\cos^2\beta} (\delta^B)^2
\,  {\tilde m}\, \epsilon_H\sim 10^{-2} eV.\nn \eea Taking
$\cos\beta\sim 1$ and $\tilde m\sim 100$ GeV, this leads to
$\delta^{\mu}\sim 10^{-6}$ and $\delta^B\sim 10^{-5}$. Here the
generation indices of the $\delta$'s are dropped for simplicity.
All these results are highly parameter space dependent, for
example in this simple-minded analysis we take $\epsilon_H\sim 1$,
but as mentioned earlier, it can be quite small. In addition, from
the diagonalization of the tree level mass matrix, one can see
that one can get large mixing angles. However to reproduce the
exact mass-squared differences and the mixing pattern, in
particular the smallish $\theta_{13}$, some fine-tuning of the
relevant parameters is necessary.

It is interesting to note that the required sizes of the
$\delta$'s are too small, whereas we expect them to be naturally
of order one. So it is interesting to explore flavor models which
can naturally explain this smallness. We already mentioned that
the approximate alignment between $\mu_{\alpha}$ and $v_{\alpha}$,
required to explain the smallness of the tree level contribution
and hence the small $\delta^{\mu}_i$'s, can be realized\cite{bgnn}
if we consider, in addition, the presence of some Abelian
horizontal symmetry\cite{LNS}. The use of an Abelian group that
distinguishes between different generations has been quite
successful in the quark sector. In Ref.~\cite{ygsr}, an
$U(1)_H$ group has been considered to get naturally small
$R$-parity violating parameters, which were then used to estimate
different contributions to the neutrino masses. However a
potential problem with this simple model is to explain a small
$\theta_{13}$ one needs a mild fine-tuning. However, it might be
possible to fit the data with less fine-tuning if one goes for an
elaborate model, like those with a more complicated flavor
symmetry group\cite{yyy}.

\section{CONCLUSION}\label{sec:conclu}

We see that $R$-parity violating minimal supersymmetric model can
naturally accommodate lepton number violation, which can generate
Majorana neutrino masses involving weak scale fields. It provides
a viable alternative to the see-saw mechanism. These models are
particularly useful to reproduce large mixing angles with a
hierarchical spectrum, as indicated by experiments. But one should
also keep in mind that, in general, to keep the tree level
contribution small enough, some fine-tuning is required.

In this $R$-parity violating model, the contributions come from
both tree and loop level diagrams, which are highly parameter
space dependent. So which of them are the most important ones
depend on the model concerned. However in a generic scenario, one
expects the tree level contributions give mass to the heaviest
neutrino. The other two neutrinos get mass at the loop level. $BB$
loops, as induced by the sneutrino--anti-sneutrino mixing, are
expected to be the dominant loop contributions. The trilinear
loops turn out to be rather small.

In addition to the constraints from neutrino masses, the $R$-parity
violating parameters suffer from tremendously small upper bounds
imposed by several flavor violating
processes\cite{rpvrev,Herbi,newrev}. It is interesting to explore
theoretical models which can naturally explain this smallness.



\section*{Acknowledgments}
The author thanks Gautam Bhattacharyya, Yuval Grossman and Amitava
Raychaudhuri for comments on the manuscript and discussions. He also
acknowledges financial support from Lady Davis Trust.


\section*{References}

\vspace*{6pt}

\begin{thebibliography}{0}

\bibitem{rev}
M.~C.~Gonzalez-Garcia and Y.~Nir,
Rev.\ Mod.\ Phys.\  {\bf 75}, 345 (2003)
[hep-ph/0202058];
Y.~Grossman,
hep-ph/0305245.

\bibitem{rev2}
G.~Altarelli and F.~Feruglio,
hep-ph/0306265;
V.~Barger, D.~Marfatia and K.~Whisnant,
Int.\ J.\ Mod.\ Phys.\ E {\bf 12}, 569 (2003)
[hep-ph/0308123].

\bibitem{SKatmo03}
M.~C.~Gonzalez-Garcia and C.~Pena-Garay,
Phys.\ Rev.\ D {\bf 68}, 093003 (2003)
[hep-ph/0306001];
M.~Maltoni, T.~Schwetz, M.~A.~Tortola and J.~W.~F.~Valle,
Phys.\ Rev.\ D {\bf 68}, 113010 (2003)
[hep-ph/0309130];
A.~Bandyopadhyay, S.~Choubey, S.~Goswami, S.~T.~Petcov and D.~P.~Roy,
Phys.\ Lett.\ B {\bf 583}, 134 (2004)
[hep-ph/0309174];
P.~C.~de Holanda and A.~Y.~Smirnov,
Astropart.\ Phys.\  {\bf 21}, 287 (2004)
[hep-ph/0309299].

\bibitem{wmap}
C.~L.~Bennett {\it et al.},
Astrophys.\ J.\ Suppl.\  {\bf 148}, 1 (2003)
[astro-ph/0302207].

\bibitem{newrev}
M.~Chemtob,
hep-ph/0406029;
R.~Barbier {\it et al.},
hep-ph/0406039.

\bibitem{valle}
J.~W.~F.~Valle,
hep-ph/9808292;
O.~C.~W.~Kong,
Int.\ J.\ Mod.\ Phys.\ A {\bf 19}, 1863 (2004)
[hep-ph/0205205].


\bibitem{Fayet:1974pd}
P.~Fayet,
Nucl.\ Phys.\ B {\bf 90}, 104 (1975);
Phys.\ Lett.\ B {\bf 69}, 489 (1977);
Phys.\ Lett.\ B {\bf 76}, 575 (1978).

\bibitem{Ibanez:1991pr}
L.~E.~Ibanez and G.~G.~Ross,
Nucl.\ Phys.\ B {\bf 368}, 3 (1992);
L.~E.~Ibanez and G.~G.~Ross,
Phys.\ Lett.\ B {\bf 260}, 291 (1991).

\bibitem{Herbi}
B.~C.~Allanach, A.~Dedes and H.~K.~Dreiner,
hep-ph/0309196.

\bibitem{ghrpv}
Y.~Grossman and H.~E.~Haber,
Phys.\ Rev.\ D {\bf 59}, 093008 (1999)
[hep-ph/9810536].

\bibitem{sachaellis}
S.~Davidson and J.~R.~Ellis,
Phys.\ Lett.\ B {\bf 390}, 210 (1997)
[hep-ph/9609451];
S.~Davidson and J.~R.~Ellis,
Phys.\ Rev.\ D {\bf 56}, 4182 (1997)
[hep-ph/9702247];
S.~Davidson,
Phys.\ Lett.\ B {\bf 439}, 63 (1998)
[hep-ph/9808425].

\bibitem{Ferr}
J.~Ferrandis,
Phys.\ Rev.\ D {\bf 60}, 095012 (1999)
[hep-ph/9810371].


\bibitem{ghbasis}
Y.~Grossman and H.~E.~Haber,
Phys.\ Rev.\ D {\bf 63}, 075011 (2001)
[hep-ph/0005276].

\bibitem{Dav-Los}
S.~Davidson and M.~Losada,
JHEP {\bf 0005}, 021 (2000)
[hep-ph/0005080];
Phys.\ Rev.\ D {\bf 65}, 075025 (2002)
[hep-ph/0010325].


\bibitem{all}
C.~S.~Aulakh and R.~N.~Mohapatra,
Phys.\ Lett.\ B {\bf 119}, 136 (1982);
L.~J.~Hall and M.~Suzuki,
Nucl.\ Phys.\ B {\bf 231}, 419 (1984);
I.~H.~Lee,
Phys.\ Lett.\ B {\bf 138}, 121 (1984);
Nucl.\ Phys.\ B {\bf 246}, 120 (1984);
G.~G.~Ross and J.~W.~Valle,
Phys.\ Lett.\ B {\bf 151}, 375 (1985);
J.~R.~Ellis, G.~Gelmini, C.~Jarlskog, G.~G.~Ross and J.~W.~Valle,
Phys.\ Lett.\ B {\bf 150}, 142 (1985);
S.~Dawson,
Nucl.\ Phys.\ B {\bf 261}, 297 (1985);
A.~Santamaria and J.~W.~Valle,
Phys.\ Lett.\ B {\bf 195}, 423 (1987);
K.~S.~Babu and R.~N.~Mohapatra,
Phys.\ Rev.\ Lett.\  {\bf 64}, 1705 (1990);
R.~Barbieri, M.~M.~Guzzo, A.~Masiero and D.~Tommasini,
Phys.\ Lett.\ B {\bf 252}, 251 (1990);
E.~Roulet and D.~Tommasini,
Phys.\ Lett.\ B {\bf 256}, 218 (1991);
K.~Enqvist, A.~Masiero and A.~Riotto,
Nucl.\ Phys.\ B {\bf 373}, 95 (1992);
J.~C.~Romao and J.~W.~Valle,
Nucl.\ Phys.\ B {\bf 381}, 87 (1992);
R.~M.~Godbole, P.~Roy and X.~Tata,
Nucl.\ Phys.\ B {\bf 401}, 67 (1993)
[hep-ph/9209251];
A.~S.~Joshipura and M.~Nowakowski,
Phys.\ Rev.\ D {\bf 51}, 2421 (1995)
[hep-ph/9408224];
Phys.\ Rev.\ D {\bf 51}, 5271 (1995)
[hep-ph/9403349];
M.~Nowakowski and A.~Pilaftsis,
Nucl.\ Phys.\ B {\bf 461}, 19 (1996)
[hep-ph/9508271];
F.~M.~Borzumati, Y.~Grossman, E.~Nardi and Y.~Nir,
Phys.\ Lett.\ B {\bf 384}, 123 (1996)
[hep-ph/9606251];
S.~Roy and B.~Mukhopadhyaya,
Phys.\ Rev.\ D {\bf 55}, 7020 (1997)
[hep-ph/9612447];
A.~S.~Joshipura, V.~Ravindran and S.~K.~Vempati,
Phys.\ Lett.\ B {\bf 451}, 98 (1999)
[hep-ph/9706482];
M.~Drees, S.~Pakvasa, X.~Tata and T.~ter Veldhuis,
Phys.\ Rev.\ D {\bf 57}, 5335 (1998)
[hep-ph/9712392];
R.~Adhikari and G.~Omanovic,
Phys.\ Rev.\ D {\bf 59}, 073003 (1999);
A.~S.~Joshipura and S.~K.~Vempati,
Phys.\ Rev.\ D {\bf 60}, 095009 (1999)
[hep-ph/9808232].
B.~Mukhopadhyaya, S.~Roy and F.~Vissani,
Phys.\ Lett.\ B {\bf 443}, 191 (1998)
[hep-ph/9808265];
K.~Choi, K.~Hwang and E.~J.~Chun,
Phys.\ Rev.\ D {\bf 60}, 031301 (1999)
[hep-ph/9811363];
S.~Rakshit, G.~Bhattacharyya and A.~Raychaudhuri,
Phys.\ Rev.\ D {\bf 59}, 091701 (1999)
[hep-ph/9811500];
D.~E.~Kaplan and A.~E.~Nelson,
JHEP {\bf 0001}, 033 (2000)
[hep-ph/9901254];
A.~S.~Joshipura and S.~K.~Vempati,
Phys.\ Rev.\ D {\bf 60}, 111303 (1999)
[hep-ph/9903435];
S.~Y.~Choi, E.~J.~Chun, S.~K.~Kang and J.~S.~Lee,
Phys.\ Rev.\ D {\bf 60}, 075002 (1999)
[hep-ph/9903465];
A.~Datta, B.~Mukhopadhyaya and S.~Roy,
Phys.\ Rev.\ D {\bf 61}, 055006 (2000)
[hep-ph/9905549];
G.~Bhattacharyya, H.~V.~Klapdor-Kleingrothaus and H.~Paes,
Phys.\ Lett.\ B {\bf 463}, 77 (1999)
[hep-ph/9907432];
A.~Abada and M.~Losada,
Nucl.\ Phys.\ B {\bf 585}, 45 (2000)
[hep-ph/9908352];
O.~Haug, J.~D.~Vergados, A.~Faessler and S.~Kovalenko,
Nucl.\ Phys.\ B {\bf 565}, 38 (2000)
[hep-ph/9909318];
E.~J.~Chun and S.~K.~Kang,
Phys.\ Rev.\ D {\bf 61}, 075012 (2000)
[hep-ph/9909429];
F.~Takayama and M.~Yamaguchi,
Phys.\ Lett.\ B {\bf 476}, 116 (2000)
[hep-ph/9910320];
R.~Kitano and K.~y.~Oda,
Phys.\ Rev.\ D {\bf 61}, 113001 (2000)
[hep-ph/9911327];
M.~Hirsch, M.~A.~Diaz, W.~Porod, J.~C.~Romao and J.~W.~Valle,
Phys.\ Rev.\ D {\bf 62}, 113008 (2000)
[Erratum-ibid.\ D {\bf 65}, 119901 (2002)]
[hep-ph/0004115];
A.~S.~Joshipura, R.~D.~Vaidya and S.~K.~Vempati,
Phys.\ Rev.\ D {\bf 62}, 093020 (2000)
[hep-ph/0006138].
J.~M.~Mira, E.~Nardi, D.~A.~Restrepo and J.~W.~Valle,
Phys.\ Lett.\ B {\bf 492}, 81 (2000)
[hep-ph/0007266];
T.~F.~Feng and X.~Q.~Li,
Phys.\ Rev.\ D {\bf 63}, 073006 (2001)
[hep-ph/0012300];
A.~S.~Joshipura, R.~D.~Vaidya and S.~K.~Vempati,
Phys.\ Rev.\ D {\bf 65}, 053018 (2002)
[hep-ph/0107204];
V.~D.~Barger, T.~Han, S.~Hesselbach and D.~Marfatia,
Phys.\ Lett.\ B {\bf 538}, 346 (2002)
[hep-ph/0108261];
A.~S.~Joshipura, R.~D.~Vaidya and S.~K.~Vempati,
Nucl.\ Phys.\ B {\bf 639}, 290 (2002)
[hep-ph/0203182];
S.~K.~Vempati,
hep-ph/0203219;
S.~K.~Kang and O.~C.~Kong,
hep-ph/0206009;
M.~A.~Diaz, M.~Hirsch, W.~Porod, J.~C.~Romao and J.~W.~F.~Valle,
Phys.\ Rev.\ D {\bf 68}, 013009 (2003)
[hep-ph/0302021].

\bibitem{ghprl}
Y.~Grossman and H.~E.~Haber,
Phys.\ Rev.\ Lett.\  {\bf 78}, 3438 (1997)
[hep-ph/9702421].

\bibitem{HKK}
M.~Hirsch, H.~V.~Klapdor-Kleingrothaus and S.~G.~Kovalenko,
Phys.\ Lett.\ B {\bf 398}, 311 (1997)
[hep-ph/9701253];
M.~Hirsch, H.~V.~Klapdor-Kleingrothaus and S.~G.~Kovalenko,
hep-ph/9701273;
M.~Hirsch, H.~V.~Klapdor-Kleingrothaus and S.~G.~Kovalenko,
Phys.\ Rev.\ D {\bf 57}, 1947 (1998)
[hep-ph/9707207];
M.~Hirsch, H.~V.~Klapdor-Kleingrothaus, S.~Kolb and S.~G.~Kovalenko,
Phys.\ Rev.\ D {\bf 57}, 2020 (1998).


\bibitem{Davidson:1999mc}
S.~Davidson, M.~Losada and N.~Rius,
Nucl.\ Phys.\ B {\bf 587}, 118 (2000)
[hep-ph/9911317].


\bibitem{biloop}
F.~Borzumati and J.~S.~Lee,
Phys.\ Rev.\ D {\bf 66}, 115012 (2002)
[hep-ph/0207184].

\bibitem{ygsr}
Y.~Grossman and S.~Rakshit,
Phys.\ Rev.\ D {\bf 69}, 093002 (2004)
[hep-ph/0311310].

\bibitem{bgnn}
T.~Banks, Y.~Grossman, E.~Nardi and Y.~Nir,
Phys.\ Rev.\ D {\bf 52}, 5319 (1995)
[hep-ph/9505248].

\bibitem{enrico}
E.~Nardi,
Phys.\ Rev.\ D {\bf 55}, 5772 (1997)
[hep-ph/9610540].

\bibitem{ghtalk}
Y.~Grossman and H.~E.~Haber,
hep-ph/9906310.

\bibitem{Haber}
H.~E.~Haber,
Nucl.\ Phys.\ Proc.\ Suppl.\  {\bf 116}, 291 (2003)
[hep-ph/0212010].

\bibitem{hunter}
J.~F.~Gunion, H.~E.~Haber, G.~L.~Kane and S.~Dawson,
``The Higgs Hunter's Guide'', Perseus Publishing, Reading, MA, 2000;
[Erratum-ibid.  hep-ph/9302272].

\bibitem{gh1}
J.~F.~Gunion and H.~E.~Haber,
Nucl.\ Phys.\ B {\bf 272}, 1 (1986)
[Erratum-ibid.\ B {\bf 402}, 567 (1993)].

\bibitem{Chun}
E.~J.~Chun, D.~W.~Jung and J.~D.~Park,
Phys.\ Lett.\ B {\bf 557}, 233 (2003)
[hep-ph/0211310].

\bibitem{numerical}
A.~Abada, S.~Davidson and M.~Losada,
Phys.\ Rev.\ D {\bf 65}, 075010 (2002)
[hep-ph/0111332];
A.~Abada, G.~Bhattacharyya and M.~Losada,
Phys.\ Rev.\ D {\bf 66}, 071701 (2002)
[hep-ph/0208009].


\bibitem{LNS}
M.~Leurer, Y.~Nir and N.~Seiberg,
Nucl.\ Phys.\ B {\bf 398}, 319 (1993)
[hep-ph/9212278];
Y.~Nir and N.~Seiberg,
Phys.\ Lett.\ B {\bf 309}, 337 (1993)
[hep-ph/9304307];
M.~Leurer, Y.~Nir and N.~Seiberg,
Nucl.\ Phys.\ B {\bf 420}, 468 (1994)
[hep-ph/9310320];
Y.~Grossman and Y.~Nir,
Nucl.\ Phys.\ B {\bf 448}, 30 (1995)
[hep-ph/9502418];
Y.~Nir and Y.~Shadmi,
hep-ph/0404113.


\bibitem{FrNi}
C.~D.~Froggatt and H.~B.~Nielsen,
Nucl.\ Phys.\ B {\bf 147}, 277 (1979).

\bibitem{yyy}
Y.~Grossman, Y.~Nir and Y.~Shadmi,
JHEP {\bf 9810}, 007 (1998)
[hep-ph/9808355];
K.~Choi, E.~J.~Chun, K.~Hwang and W.~Y.~Song,
Phys.\ Rev.\ D {\bf 64}, 113013 (2001)
[hep-ph/0107083].

\bibitem{rpvrev} For reviews, see for example,
G.~Bhattacharyya,
Nucl.\ Phys.\ Proc.\ Suppl.\  {\bf 52A}, 83 (1997)
[hep-ph/9608415];
H.~K.~Dreiner,
hep-ph/9707435;
G.~Bhattacharyya,
hep-ph/9709395;
R.~Barbier {\it et al.},
hep-ph/9810232;
S.~Raychaudhuri,
hep-ph/9905576;
B.~C.~Allanach, A.~Dedes and H.~K.~Dreiner,
Phys.\ Rev.\ D {\bf 60}, 075014 (1999)
[hep-ph/9906209].



\end{thebibliography}
\end{document}